\documentclass[a4paper,fleqn]{cas-sc}

\usepackage[authoryear]{natbib}
\usepackage{graphicx,times}
\usepackage{subcaption}
\usepackage{amssymb,amsmath}
\usepackage{aas_macros}
\usepackage{hyperref}
\usepackage{comment}
\let\oldAA\AA
\renewcommand{\AA}{\text{\normalfont\oldAA}}
\usepackage{pdflscape}

\def\tsc#1{\csdef{#1}{\textsc{\lowercase{#1}}\xspace}}
\tsc{WGM}
\tsc{QE}
\begin{document}

\let\WriteBookmarks\relax
\def\floatpagepagefraction{1}
\def\textpagefraction{.001}

\shorttitle{X-ray polarization of PKS 2155-304 and 3C 454.3}    

\shortauthors{A M Bharathan \& C S Stalin\& S Sahayanathan  \& B Mathew}  

\title[mode = title]{Clues on the X-ray emission mechanism of blazars PKS 2155\textendash 304 and 3C 454.3 through polarization studies}
\tnotemark[<tnote number>] 

%\tnotetext[<tnote number>]{<tnote text>} 


\author[1]{Athira M Bharathan}

\cormark[1]

%\fnmark[<footnote mark no>]

\ead{athirabharathan1997@gmail.com.com}
\affiliation[1]{organization={Department of Physics and Electronics, CHRIST (Deemed to be University)},
            city={Bangalore},
%          citysep={}, % Uncomment if no comma needed between city and postcode
            postcode={560029}, 
            state={Karnataka},
            country={India}}
            
\author[2]{C S Stalin}

% Footnote of the second author
%\fnmark[]

% Email id of the second author
\ead{stalin@iiap.res.in}

\affiliation[2]{organization={Indian Institute of Astrophysics, Block II, Koramangala},
            city={Bangalore},
%          citysep={}, % Uncomment if no comma needed between city and postcode
            postcode={560029}, 
            state={Karnataka},
            country={India}}

\author[3,4]{Sunder Sahayanathan}

% Footnote of the second author
%\fnmark[]

% Email id of the second author
\ead{sunder@barc.gov.in}

\affiliation[3]{organization={Astrophysical Sciences Division, Bhabha Atomic Research Centre},
            city={Mumbai},
%          citysep={}, % Uncomment if no comma needed between city and postcode
            postcode={400085}, 
            state={Maharashtra},
            country={India}}
            
 \affiliation[4]{organization={Homi Bhabha National Institute},
            city={Mumbai},
%          citysep={}, % Uncomment if no comma needed between city and postcode
            postcode={400094}, 
            state={Maharashtra},
            country={India}}

\author[1]{Blesson Mathew}

% Footnote of the second author
%\fnmark[]

% Email id of the second author
\ead{blesson.mathew@christuniversity.in}

%\affiliation[1]{organization={Department of Physics and Electronics, CHRIST (Deemed to be University)},
 %           city={Bangalore},
%          citysep={}, % Uncomment if no comma needed between city and postcode
 %           postcode={560029}, 
 %           state={Karnataka},
 %           country={India}}

%\ead[url]{<URL>}

%\credit{<Credit authorship details>}

%\affiliation[1]{organization={Department of Physics, Farook College},
%            city={Calicut},
%          citysep={}, % Uncomment if no comma needed between city and postcode
%            postcode={673632}, 
 %           state={Kerala},
 %           country={India}}

\credit{}

%\affiliation[<aff no>]{organization={},
%          citysep={}, % Uncomment if no comma needed between city and postcode
 %           postcode={}, 
 %           state={},
 %           country={}}

\cortext[1]{Corresponding author}


\begin{abstract}
X-ray polarization measurable with the imaging X-ray Polarimetry Explorer (\textit{IXPE}) could constrain the long debated leptonic versus hadronic  origin for the high energy component in the broad band spectral energy distribution (SED) of blazars. We report here the results from \textit{IXPE} and SED modeling of PKS 2155$-$304 and 3C 454.3, a high and low synchrotron peaked blazar. For PKS 2155$-$304, from model-independent analysis, we found  polarization angle $\Psi_X$ = (130 $\pm$ 2.5) deg and polarization degree $\Pi_X$ = (20.9 $\pm$ 1.8)\% in the 2$-$8 keV band in agreement with  spectro-polarimetric analysis.  We found $\Pi_X$ to vary with time and indications of it to vary between energies, suggesting that the emission regions are stratified.  For 3C 454.3, we did not detect X-ray polarization in the June 2023 observation, analyzed here for the first time. The detection of X-ray polarization in PKS 2155$-$304 and its non-detection in 3C 454.3 is in accordance with the X-ray emission from synchrotron and inverse Compton process, respectively, operating in these sources. Further, our division of the dataset into finer time bins allows a more granular view of polarization variability. Additionally, we modeled the broadband SEDs of both the sources using data acquired quasi-simultaneously with \textit{IXPE}, in the optical, UV and X-rays from {\it Swift}, {\it AstroSat} and $\gamma$-rays from {\it Fermi}. In PKS 2155$-$304, the observed X-ray is found to lie in the high energy tail of the synchrotron component of the SED, while in 3C 454.3 the observed X-ray lies in the rising part of the inverse Compton component of the SED. Our SED modeling along with X-ray polarization observations favour a leptonic scenario for the observed X-ray emission in PKS 2155$-$304.  The SED modeling for these specific \textit{IXPE} epochs has not been presented before, allowing us to place additional constraints on the physical conditions in the jet. These results strengthen the case for a structured jet model where X-ray emission originates from a compact acceleration zone near the shock front, while lower-energy optical emission is produced in a broader, more turbulent region.

\end{abstract}

\begin{keywords}
techniques: polarization \sep galaxies: active \sep BL Lacertae objects:individual:PKS 2155-304 - quasars: individual 3C 454.3 \sep X-rays: galaxies

\end{keywords}

\maketitle
\section{Introduction}
\label{sec1}
Blazars are a subclass of active galactic nuclei (AGN), the exceptionally bright objects in the Universe, with luminosities ranging from 10$^{42}$ to 10$^{48}$ erg s$^{-1}$ and believed to be powered by accretion of matter onto supermassive black holes (M$_{BH}$ = 10$^6$ to 10$^{10}$ M$_{\odot}$) at the centres of galaxies \citep{1969Natur.223..690L,1973A&A....24..337S}.  Their radiation output is dominated by non-thermal processes in their relativistic jets, oriented close to the line of sight to the observer. Blazars emit radiation throughout the electromagnetic spectrum from low-energy radio frequencies to high-energy TeV $\gamma$-rays \citep{2019NewAR..8701541H}. Their radiation output at all wavelengths show variations on time scales varying from minutes to years \citep{1995ARA&A..33..163W,1997ARA&A..35..445U}. In addition to flux variations, they are known to be highly polarized in the optical \citep{1980ARA&A..18..321A} and X-ray wavelengths \citep{2024JApA...45...35B,2024ApJ...975..185B}, and also show optical \citep{2005A&A...442...97A,2017ApJ...835..275R} and X-ray polarization variations \citep{2022Natur.611..677L,2022ApJ...938L...7D, 2024JApA...45...35B, 2024ApJ...963L..41H, 2024A&A...689A.119K}.

Blazars are sub-divided into two groups, namely flat spectrum radio quasars (FSRQs) and BL Lac objects (BL Lacs), characterized by the presence of strong (EW $>$ 5\,{\AA})  and weak (EW $<$ 5\,{\AA}) or no emission lines respectively in their optical spectra \citep{1991ApJ...374..431S}. Their broadband spectral energy distribution (SED) shows a double component structure. The synchrotron emission mechanism, which occurs when relativistic electrons spiral around the magnetic fields within their relativistic jet, is responsible for the low-energy component of their SED, and peaks between the UV and X-ray frequencies. The high energy component of their SED peaks in the GeV to TeV energy range, and the physical process that gives rise to the high energy component is uncertain and debated \citep{2013ApJ...768...54B, 2018ApJ...863...98P, 2019Galax...7...20B}. Depending on the synchrotron peak in their SEDs, blazars are further sub-divided in low synchrotron peaked (LSP; $\nu_{peak}$ $<$ 10$^{14}$ Hz), intermediate synchrotron peaked (ISP; 10$^{14}$ Hz $<$ $\nu_{peak}$ $<$ 10$^{15}$ Hz) and high synchrotron peaked (HSP; $\nu_{peak}$ $>$ 10$^{15}$ Hz) blazars, respectively \citep{2010ApJ...716...30A}. Most of the FSRQs belong to the LSP blazar category, while a large fraction of HSP blazars are BL Lac objects \citep{2014MNRAS.438.3058R,2020A&A...634A..80R}.

Both leptonic and hadronic models are proposed to explain the high energy emission process in blazars (\citealt{2013ApJ...768...54B,2019Galax...7...20B} and references therein). In the leptonic scenario, the high energy emission is because of inverse Compton scattering, wherein the synchrotron emitting electrons in the jet Compton up scatter either the synchrotron photons (a process called synchrotron self Compton; SSC e.g., \citealt{1992ApJ...397L...5M}) or photons from sources external to the jet (a process called external Compton; EC e.g., \citealt{1992A&A...256L..27D,1994ApJ...421..153S}). In the hadronic scenario, the high energy emission could be due to proton synchrotron process or due to secondary particles in photo-pion cascades \citep{1989A&A...221..211M}. Though hadronic models are generally less favoured owing to the large energy requirements \citep{2013ApJ...768...54B}, the recent detection of neutrinos associated with the blazar TSX 0506+057 \citep{2018Sci...361.1378I}, and other handful of blazars \citep{2020A&A...640L...4G,2023MNRAS.519.1396S} suggests that hadronic processes too may be present in blazar jets, which is also hinted in the broadband SED analysis of the neutrino blazar PKS 0735+178 in its faint state \citep{2024MNRAS.529.3503B}.

Different particle acceleration and emission mechanisms in blazar jets lead to inherent differences in the measured polarization across wavelengths. Therefore, multi-wavelength polarization observations are important diagnostic tools to constrain emission and acceleration processes as well as the magnetic field strengths in blazar jets \citep{2013ApJ...774...18Z,2024ApJ...967...93Z}. As the X-ray emission in HSP and LSP blazars lies respectively in the electron synchrotron and the high energy inverse Compton part of their SEDs, a comparative analysis of the X-ray polarization of LSP and HSP blazars can provide vital clues to the leptonic versus hadronic origin of X-ray emission from blazar jets. For instance, polarized X-ray emission from LSP blazars can favour proton synchrotron model. X-ray polarization measurements available till now have detected significant X-ray polarization in HSP blazars, while only upper limits to X-ray polarization are known in LSP blazars \citep{2024A&A...689A.119K,2024ApJ...972...74M}.  These limited observations must not be taken as proof to exclude hadronic processes in blazar jets. A support to this comes from the requirement of a hadronic component to explain the faint state SED of a neutrino blazar PKS 0735+178 \citep{2024MNRAS.529.3503B}. In this paper we report our analysis of X-ray polarization as well as SED modeling of PKS 2155$-$304 (a HSP blazar) and 3C 454.3 (a LSP blazar). This paper is structured as follows: Section 2 describes the observation data and its reduction. The analysis, results, and broadband SED fitting are presented in Section 3. The results are discussed in Section 4 followed by the summary in the final Section.

\begin{figure}
\centering
\includegraphics[width=0.45\textwidth]{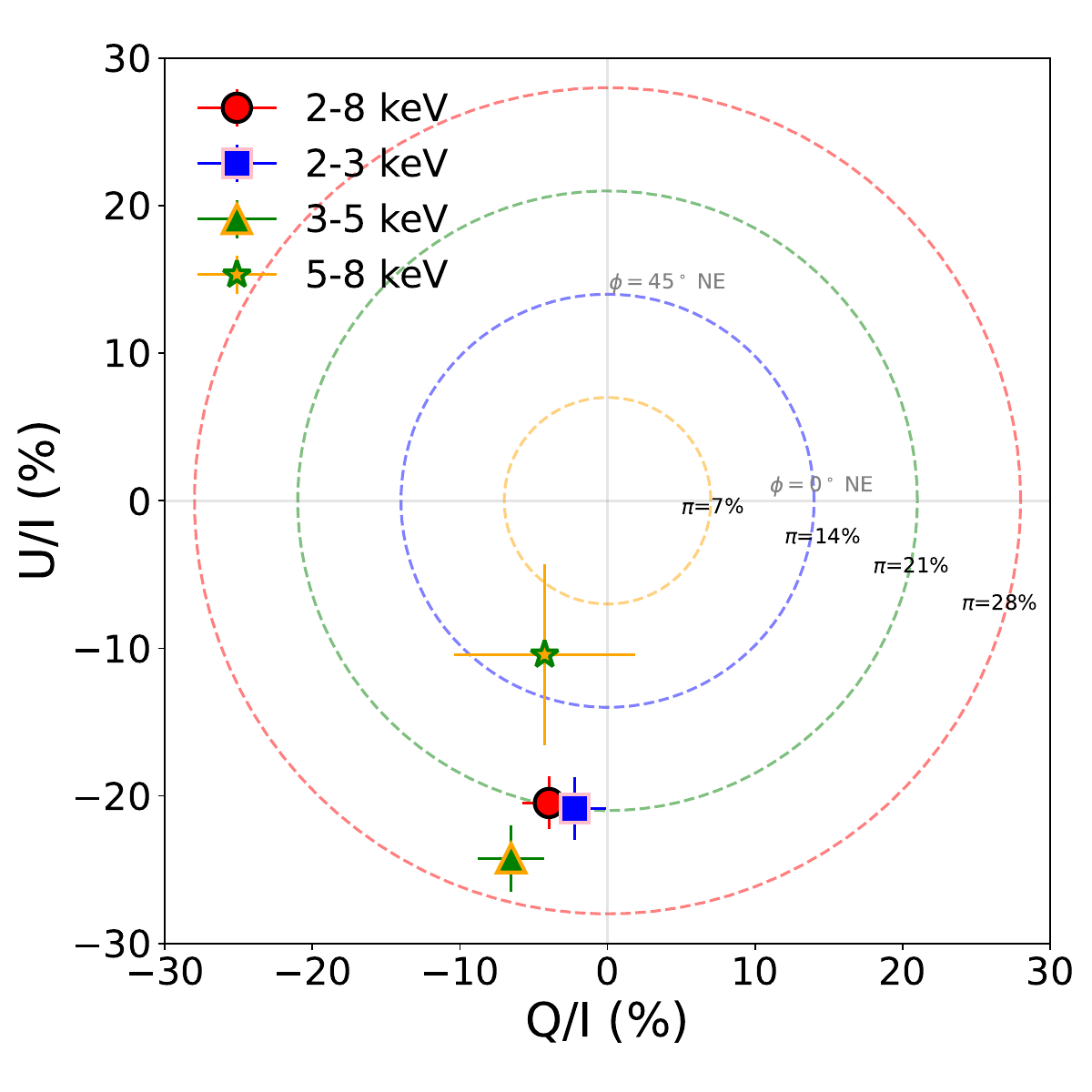}
\caption{Normalized U/I and Q/I Stokes parameter in the total 
2$-$8 keV (red circle), 2$-$3 keV (blue square), 3$-$5 keV (green triangle) and 5$-$8 keV (orange star ) band of
{\it IXPE} on 27 October 2023 for the source PKS 2155$-$304.}
\label{figure-1}
\end{figure}

\section{Observations and data reduction}\label{sec2}

For X-ray polarization measurements, we used data from the \textit{Imaging X-ray Polarimetry Explorer }(\textit{IXPE}) while for SED modeling we used data in the $\gamma$-rays from {\it Fermi}, X-rays from {\it Swift}-XRT, UV and optical bands from {\it Swift}-UVOT and {\it AstroSat}-UVIT that were acquired quasi-simultaneously with \textit{IXPE} observations. The details of the data used and the associated data reduction are summarized below.
\label{sec:obs}
\subsection{\textit{Fermi}-LAT}
To reduce the data from the Large Area Telescope (LAT; \citealt{2009ApJ...697.1071A}) onboard {\it Fermi}, we used the {\it fermipy} package \citep{2017ICRC...35..824W}. We analyzed data in the energy range from 100 MeV to 300 GeV, considering all SOURCE class events (evclas=128 and evtype=3) within a region of interest of 15 deg centered on the source. To ensure data quality, we applied the filter "DATA\_QUAL $>$ 0 \&\& LAT\_CONFIG==1".  For the background models, we used the galactic diffuse emission model (gll\_iem\_v06) and the isotropic diffuse emission model (iso\_P8R2\_SOURCE\_V6\_v06). We performed a binned likelihood analysis to generate the $\gamma$-ray spectra. For PKS 2155$-$304, we used data covering the period between 07 October 2023 (MJD 60244) and 16 November 2023 (MJD 60264). For 3C 454.3, we used the data spanning the period between 10 June 2023 and 30 June 2023.

\subsection{\textit{IXPE}}
\textit{IXPE} observed PKS 2155$-$304 on October 27, 2023, and 3C 454.3 on June 20, 2023, using its three detector units (DU) with a total exposure time of 476 ksec and 247 ksec, respectively. The log of  \textit{IXPE} observation is given in Table \ref{table-1}. We utilized the cleaned and calibrated level 2 data and used the version 30.0.0 of the {\it ixpeobssim} software \citep{2022SoftX..1901194B} to analyze the data. To create a count map in sky coordinates, we employed the CMAP algorithm within the {\it xpbin} task, which is the main interface for binning the photons. To extract the source counts we used a circular region centred on the source with a 60$^{\prime\prime}$ radius adopted across the three DUs, while to extract the background counts we used a source-free region with a 100$^{\prime\prime}$ radius for each DU. We used the {\it xpselect} task  to generate filtered source and background regions for subsequent polarimetric analysis. For spectro-polarimetric analysis, we generated I, Q, and U source and background spectra using the PHA1I, PHA1Q, and PHA1U algorithms within the {\it xpbin} task of {\it ixpeobssim} for the three DUs.

\subsection{Swift-XRT}
We analyzed X-ray data for the blazars PKS 2155$-$304 and 3C 454.3  using observations from the  X-Ray Telescope (XRT; \citealt{2005SSRv..120..165B}), onboard the Neil Gehrels {\it Swift} Observatory \citep{2004ApJ...611.1005G} which we accessed from the HEASARC archives\footnote{https://heasarc.gsfc.nasa.gov/cgi-bin/W3Browse/w3browse.pl}. The X-ray data covered an energy range of 0.3 to 10 keV. For PKS 2155$-$304, we used the two epochs of observations carried out on 27 October (OBSIDs = 00097172005, 00097172006) coinciding with the day of \textit{IXPE} observations. For 3C 454.3, we used the data acquired by {\it Swift} on 19 June 2023 (OBSID = 00097167003), about a day prior to \textit{IXPE} observations. 

We performed the data reduction using the default parameters recommended by the instrument team. We extracted the source spectra from a circular region centred on the source with a radius of 60$^{\prime\prime}$ for PKS 2155$-$304 and 20$^{\prime\prime}$ for 3C 454.3, while for the background spectra we used a  circular region away from the source with a radius of 120$^{\prime\prime}$ and 50$^{\prime\prime}$ for PKS 2155$-$304 and 3C 454.3, respectively. We combined the exposure map using the XIMAGE tool and generated the ancillary response files with the {\it xrtmkarf} tool. We carried out spectral fitting using an absorbed simple power-law model, incorporating a weighted average Galactic neutral hydrogen column density of 1.29 $\times$ 10$^{20}$ cm$^{-2}$ for PKS 2155$-$304 in XSPEC \citep{1999ascl.soft10005A}. For 3C 454.3 we used average Galactic neutral hydrogen column density of 6.78 $\times$ 10$^{20}$ cm$^{-2}$.

\subsection{\textit{Swift}-UVOT}
For both PKS 2155$-$304 and 3C 454.3, we utilized UV and optical data from the \textit{Swift} Ultraviolet/Optical Telescope (UVOT; 1600$-$8000\,{\AA}; \citealt{2005SSRv..120...95R}), an instrument onboard the {\it Swift} spacecraft. For PKS 2155$-$304, we  used the data in U filter observed on 27 October 2023 (OBSID = 00097172005) and in U and W2 filters observed on the same day with OBSID of 00097172006. Similarly, for 3C 454.3 we used the data acquired in V, U, W1 and W2 filters observed on 19 June 2023 with the OBSID of 00097167003. We processed the data using the online tool provided by the telescope’s data archive. Here, photometry was carried out using a circular aperture of 5 arcsec radius and the background was extracted using an annular region centered on the source with an inner and outer radii of 27.5 and 35 arcsec, respectively.
The measurements were corrected for Galactic extinction with E(B$-$V) value of 0.019 for PKS 2155$-$304 and 0.093 for 3C 454.3.

\subsection{\textit{AstroSat}-UVIT}
3C 454.3, in addition to {\it Swift}-UVOT was also observed on 26 June 2023, in the UV using the Ultra-Violet Imaging Telescope \citep{2017CSci..113..583T} onboard {\it AstroSat} \citep{2017JApA...38...27A}. It was launched by the Indian Space Research Organization on 28 September 2015 as India's multi-wavelength astronomical observatory. These observations (OBSID = T05\_124T01\_9000005712)  were carried out in three FUV filters, namely F154W (BaF2), F1172M (Silica) and F169M (Sapphire) with central wavelengths of 1541\,{\AA}, 1717\,{\AA} and 1608\,{\AA} , respectively. For the analysis of data from UVIT, we obtained the science ready Level 2 data directly from the Indian Space Science Data Center (ISSDC)\footnote{https://www.issdc.gov.in/astro.html}. The science ready images were made available to ISSDC by the UVIT Payload Operations Center (POC) at the Indian Institute of Astrophysics, Bangalore. At the POC, the Level 1 data from UVIT were processed using the UVIT L2 pipeline version 6.3 \citep{2021JApA...42...29G,2022JApA...43...77G}. The pipeline corrects for the drift of the spacecraft, flat field and geometric distortion. We carried out aperture photometry using a radius of 12 sub-pixels. We removed the background, estimated from a circular annular regions in the source with an inner and outer radius of 25 and 35 sub-pixels, respectively. The brightness derived over a radius of 12 sub-pixel radius was later converted to the total brightness of the source using the correction given in \cite{2020AJ....159..158T}. We then converted the measured brightness into flux units using the unit conversion factors given in \cite{2017AJ....154..128T}. These measurements were also corrected for Galactic extinction. Details of the multi-wavelength observations used in this work are given in Table \ref{table-2}.

\begin{table}
%\label{table-1}
\centering
\caption{\small{The log of {\it IXPE} observations. The details are the observational ID (OBSID), 
date of observation and the exposure time.}}\label{table-1}
\begin{tabular}{lcll}
\hline
Source & OBSID & Date & Exposure time (s) \\
\hline 
PKS 2155$-$304 & 02005601 & 27 October 2023 & 476108 \\
3C 454.3       & 02008901 & 20 June 2023    & 274721 \\
\hline 
\end{tabular}
\end{table}

\begin{table}
\centering
\caption{Details of the multi-wavelength data used in this work.}
\label{table-2}
\resizebox{\textwidth}{!}{%
\begin{tabular}{lll} 
\hline 
Telescope     & \multicolumn{2}{c}{Name of the source} \\ 
\cline{2-3}
              & ~~~~~~~~~~PKS 2155$-$304                          &   ~~~~~~~~~~3C 454.3  \\ 
\hline
\textit{Fermi}         & 07 October 2023 $-$ 16 November 2023               &  10 June 2023 $-$ 30 June 2023  \\
\textit{Swift}/XRT     & 27 October 2023; OBSID = 00097172005              &  19 June 2023; OBSID = 00097167003     \\
                       & 27 October 2023; OBSID = 00097172006              &                                       \\
\textit{Swift}/UVOT    & 27 October 2023, OBSID = 00097172005; Filters = U &  19 June 2023; OBSID = 00097167003; Filters = V, U, W1, W2   \\
                       & 27 October 2023; OBSID = 00097172006; Filters = U, W2  &                               \\
\textit{AstroSat}/UVIT &                                                   &  27 June 2023; OBSID = T05\_124T01\_9000005712 ;  \\
                       &                                                   & Filters = F154W, F1172M and F169M \\ 
\hline
\end{tabular}%
}
\end{table}

\section{Analysis and Results}
\label{sec:anl}
\subsection{Polarimetry}
Using the {\tt PCUBE} method in the {\tt xpbin} task, we examined the polarimetric signal from PKS 2155$-$304 and 3C 454.3. For each of the three DUs, we created a polarization cube in order to extract information such as the minimum detectable polarization (MDP), the polarization angle ($\Psi_{X}$), and the polarization degree ($\Pi_{X}$) along with their corresponding errors. We calculated $\Psi_{X}$ and $\Pi_{X}$ using the Stokes parameters $Q$ and $U$. The normalized Stokes parameters are expressed as, $q = Q/I$ and $u = U/I$. The $\Pi_{X}$ is calculated as,
\begin{equation}
\Pi_{X} = \frac{\sqrt{Q^2+U^2}}{I}
\end{equation}
while $\Psi_{X}$ is calculated as,
\begin{equation}
\Psi_X = \frac{1}{2} \arctan \left(\frac{u}{q}\right)
\end{equation}

We detected significant polarization in the 2$-8$ keV band in  PKS 2155$-$304 but not for 3C 454.3. The source 3C 454.3 was first observed  with \textit{IXPE} on 31 May 2022 (OBSID = 01005401). During that epoch too, only the upper limit to the X-ray polarization parameters were obtained. This non-detection in X-ray was also coincident with its low ($<$ 1\%) optical polarization during that epoch \citep{2024ApJ...972...74M}.  For PKS 2155$-$304, in the 2$-$8 keV band using the complete dataset of 476 ksec, we obtained $\Pi_X$ = (20.9 $\pm$ 1.8)\%, $\Psi_X$ = (130.0 $\pm$ 2.5) deg. This is also in agreement with the values of  $\Pi_X$ = (21.9 $\pm$ 1.9)\%, $\Psi_X$ = (129.9 $\pm$ 2.5) deg reported by \cite{2024ApJ...963L..41H}. Also, using the same data set and dividing the \textit{IXPE} pointing of 476 ksec into two equal time bins, \cite{2024A&A...689A.119K} found \textbf{$\Psi_X$} to be stable within errors with values of $\Psi_X$ = (192.4 $\pm$ 1.8) deg and (125.4 $\pm$ 3.9) deg for the first and second time bin, respectively, while $\Pi_X$ was found to be variable with values of $\Pi_X$ = (30.7 $\pm$ 2.0)\% for the fist time bin and $\Pi_X$ = (15.3 $\pm$ 2.1)\% for the second time bin, respectively. As polarization was detected in PKS 2155$-$304 in the 2$-$8 keV band,  we also checked for energy dependence of polarization in it. For this, we divided the \textit{IXPE} energy range of 2$-$8 keV into three energy bins, namely 2$-$3 keV, 3$-$5 keV and  5$-$8 keV. We found polarization to be significantly detected in the 2$-$3 keV and 3$-$5 keV bands. However, in the 5$-$8 keV band, the derived polarization degree $\Pi_X$ = (11.3 $\pm$ 6.9)\% is below the corresponding MDP value of 18.6\%. This does not represent a significant detection but rather the best-fit estimate returned by the PCUBE method along with its statistical uncertainty. Within the quoted errors, however, the value approaches the MDP threshold, suggesting that the source could possess a polarization level near the detectability limit, though not significant at the 99\% confidence level. Therefore, the quoted polarization in the 5$-$8 keV band in PKS 2155-304 should be considered as upper limit, rather than firm detection. A similar situation applies to 3C 454.3, where in the 2$-$8 keV band the measured $\Pi_X$ = (3.7 $\pm$ 4.3)\% lies below the MDP of 13\%, and thus undetected.
%but not in the 5$-$8 keV band for which we obtained a value of $\Pi_X$ = (11.3 $\pm$ 6.9)\% which is much lower than the MDP value of 18.6\%. 
The results of polarization measurements are given in Table \ref{table-3}. From Table \ref{table-3} it is clear that, there is evidence for energy dependent polarization in PKS 2155$-$304 with  an  increase in  $\Pi_{X}$ from 2$-$3 keV to  3$-$5 keV. \cite{2024ApJ...963L..41H} too found evidence for an energy dependent polarization with $\Pi_X$ = (21.0 $\pm$ 2.1)\% in the 2$-$3 keV band, increasing to $\Pi_X$ = (28.6 $\pm$ 2.7)\% in the 3$-$4  keV band and no polarization detected above the 5 keV band. Dividing the 2$-$8 keV energy range to many smaller time bins  \cite{2024A&A...689A.119K} found no difference in the polarization properties between different energy bins.  
The normalized U/I and Q/I Stokes parameters obtained from the combined cube is illustrated in Fig.\ref{figure-1}. 

\begin{figure}
    \centering
     \includegraphics[scale=0.35]{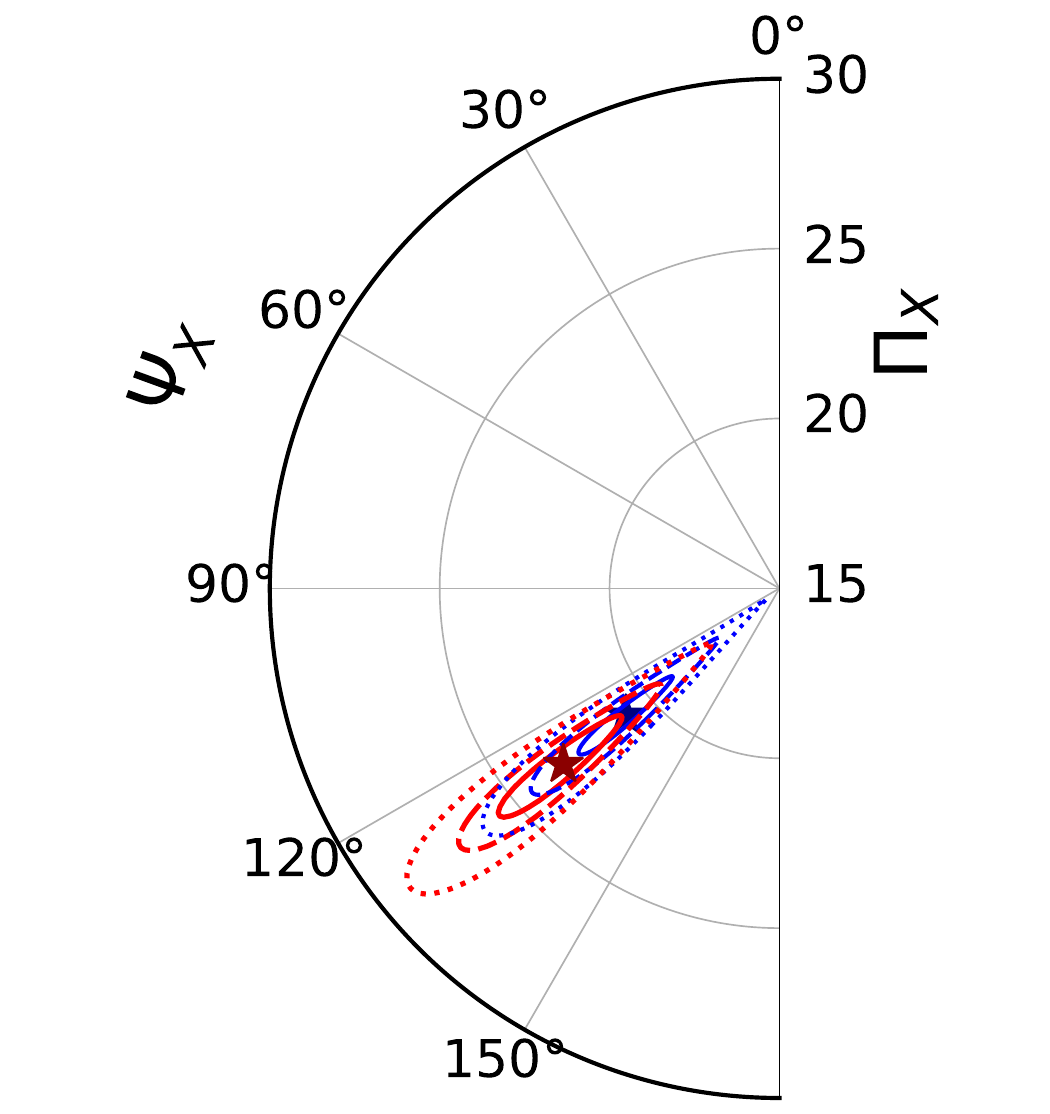}
    \caption{The position of the measured polarization in the 2$-$8 keV band for PKS 2155$-$304 from model independent PCUBE analyses (blue) and 
spectro-polarimetric fitting in XSPEC (red). Here,  $\Pi_{X}$ is given in percentage and $\Psi_{X}$ is given in degrees. The 
contours represent the 1$\sigma$, 2$\sigma$ and 3$\sigma$ uncertainties.}
    \label{figure-2}
\end{figure}

\subsection{Spectro-polarimetry}
In addition to the model independent approach, we also carried out spectro-polarimetic analysis to determine the polarization properties. This was done only for PKS 2155$-$304, as it was detected in X-ray polarization from model independent approach. We carried out the spectro-polarimetric fit in XSPEC \citep{1999ascl.soft10005A}. We modeled the spectra using a  {\it power law}, multiplied by an energy independent polarization term {\it polconst}. The model  has the following form in XSPEC,
\begin{equation}
 constant \times TBabs \times (polconst \times pow)
\end{equation}
Here, the {\it constant} represents the inter-calibration constant for each detector. {\it TBabs} was used to model the Milky Way Galactic hydrogen column density, which was taken from \cite{2013MNRAS.431..394W}. During the fit, the column density ($N_{H}$) was fixed. This model provided a good fit to the I, Q, and U spectra (from the three detectors). %Fig.\ref{figure-1} displays the best fit I, Q, and U spectra along the residuals.  
The errors associated with the model parameters were calculated at the 90$\%$ confidence level using the $\chi^2$ = 2.71 criterion. The resultant best-fit parameters are given in Table \ref{table-4}. We found $\Pi_X$ = (23.3 $\pm$ 2.5)\% and $\Psi_X$ = (129 $\pm$ 3.1) deg, which is in agreement with the values of $\Pi_X$ = (20.9 $\pm$ 1.8)\% and $\Psi_X$ = (130.0 $\pm$ 2.5) deg found from the model independent PCUBE analysis. For comparison the values of $\Pi_X$ and  $\Psi_X$ obtained from model independent PCUBE analysis and the model dependent spectro-polarimetric analysis are shown in Fig. \ref{figure-2}. For the same data set using a log parabola model in the spectro-polarimetric fit
\cite{2024A&A...689A.119K} found $\Pi_X$ = (23.3 $\pm$ 1.5)\%  and $\Psi_X$ = (128 $\pm$ 1.8) deg. Our results using power law model fit is also in agreement with that of 
\cite{2024A&A...689A.119K}.

\begin{table}
\centering
\caption{The measured polarization parameters in different energy bands. Given in parenthesis 
against the $\Pi_X$ values in each energy range are the MDP values.}
\label{table-3}
\resizebox{\textwidth}{!}{%
\begin{tabular}{lccccccc}
\hline
Source & OBSID  & \multicolumn{3}{c}{$\Pi_X$ (\%)} & \multicolumn{3}{c}{$\Psi_X$ (degrees)} \\
\hline
       &        & 2$-$3 keV          & 3$-$5 keV          & 2$-$8 keV          & 2$-$3 keV   & 3$-$5 keV   & 2$-$8 keV \\
\hline
PKS 2155$-$304 & 02005601  & 20.9$\pm$2.1 (6.4) & 25.1$\pm$2.3 (6.9) & 20.9$\pm$1.8 (5.4) & 132.0$\pm$2.9 & 128.0$\pm$2.6 & 130.0$\pm$2.5 \\
3C 454.3       & 02008901  & -                  & -                  & 3.7$\pm$4.3 (13)   & -             & -             & 101$\pm$34 \\
\hline
\end{tabular}%
}
\end{table}

\begin{table}
\caption{Best fit parameters for PKS 2155$-$304 from spectro-polarimetric model fit in XSPEC}
\label{table-4}
\begin{tabular}{ccc} \hline
Model component  & Parameter  & Value   \\ \hline
%constant         &            &         \\ 
TB$_{abs}$       &  N$_H$     & (fixed) \\
\textit{polconst}         & $\Pi_X$    &  (23.3 $\pm$ 2.5)\%        \\
                 & $\Psi_X$   &  (129 $\pm$ 3.1) deg       \\
pow              & $s$   &  2.95   \\ \hline
Goodness of fit  & $\chi^2$/dof & 1.1  \\ \hline

\end{tabular}
\end{table}

\subsection{Polarization variability}
\begin{figure}
\centering
\hspace*{-0.5cm}\includegraphics[scale=0.38]{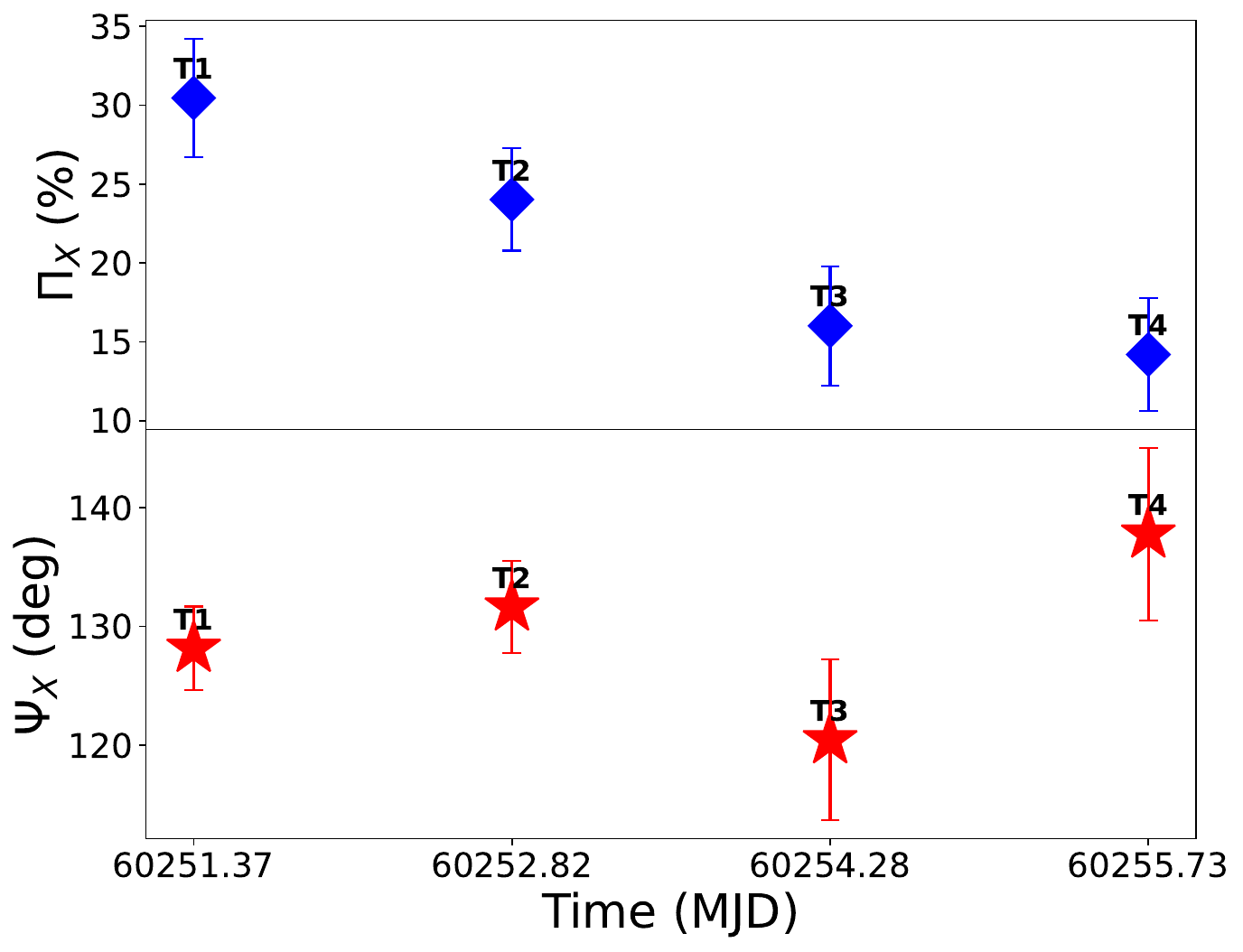}
\caption{Results of polarization variation analysis in the 2$-$8 keV band as a function of time for the source PKS 2155$-$304.}
\label{figure-3}
\end{figure}

\begin{figure}
\centering
\hspace*{-0.5cm}\includegraphics[scale=0.45]{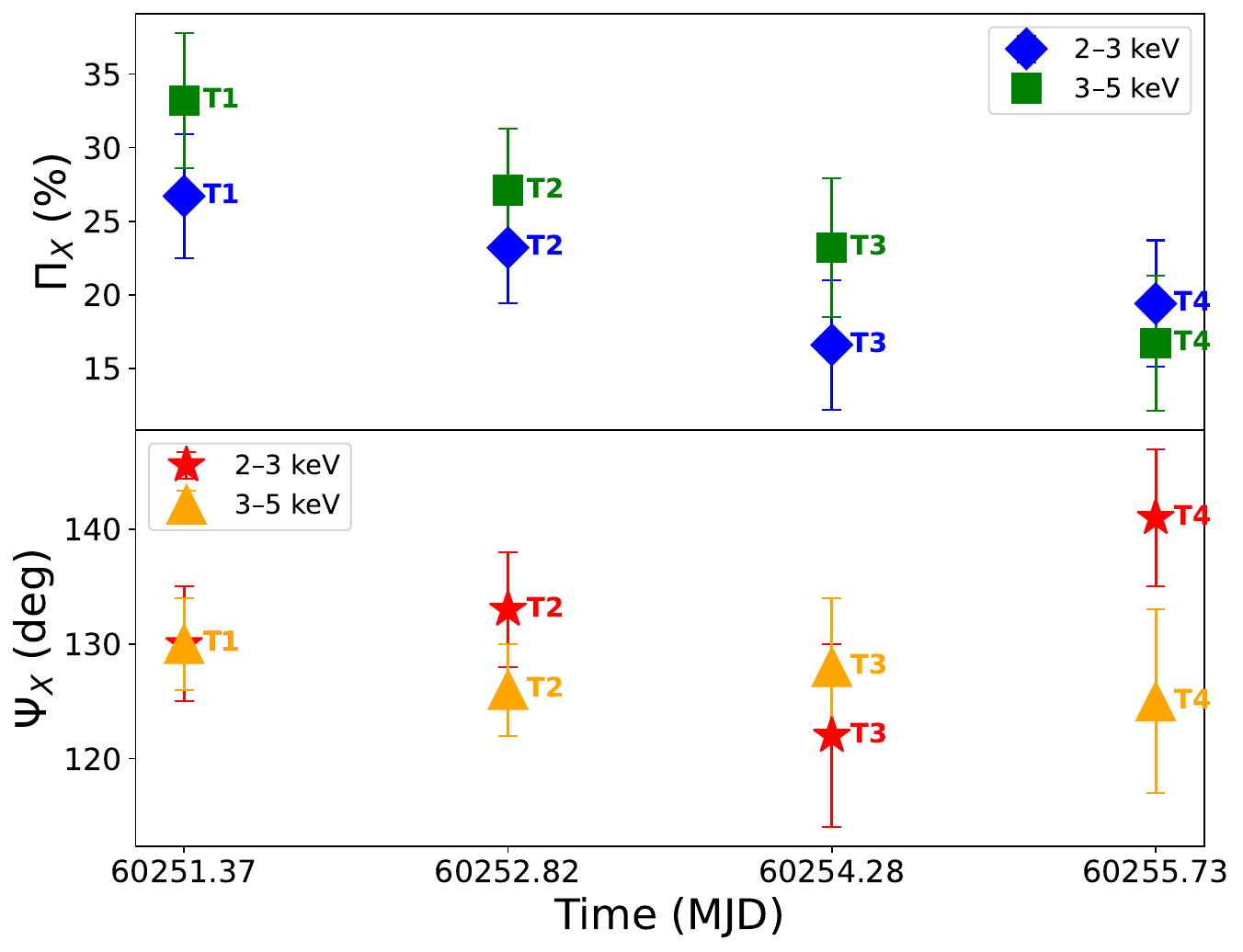}
\caption{Results of polarization variation analysis in the 2$-$3 and 3$-$5 keV band as a function of time for the source PKS 2155$-$304.}
\label{figure-235}
\end{figure}

For PKS 2155$-$304, we investigated if the observed X-ray polarization varies with time. For this we divided the whole time range in four equal bins and determined the polarization in the 2$-$8 keV band using the model independent PCUBE method. The variation in polarization as a function of time is shown in Fig. \ref{figure-3} and the results are given in Table \ref{table-5}. To check for variations if any in the measured  X-ray polarization, we carried out a $\chi^2$ test as
\begin{equation}
\chi^2 = \sum_{i = 1}^{N} \frac{(\Pi_{X_i} - \overline{\Pi}_X)^2}{\Pi_{X_i,err}^2}
\end{equation}
Here, $\Pi_{X_i}$ is the degree of polarization in the i$^{th}$ time bin, $\overline{\Pi}_X$ is the average of the N measurements and $\Pi_{X_i,err}$ is the error associated with each measurement. We found the $\chi^2$ to exceed the critical value at the 99\% confidence level. We thus have found variation in the X-ray polarization in PKS 2155$-$304 in 2$-$8 keV band. This is in agreement with the variations in $\Pi_X$ observed between two equally separated time bins from the same data set \citep{2024A&A...689A.119K}. We also performed variability analysis in the 2$-$3 keV and 3$-$5 keV bands, where polarization detections were made. Though the errors are large, variations of $\Psi$ with time is seen in both the 2$-$3 keV and 3$-$5 keV bands. The results are given in Table \ref{table-235} and the variations are shown in Fig. \ref{figure-235}.

\begin{table}
\caption {X-ray polarization variability in PKS 2155$-$304 in 2$-$8 keV band.}
\label{table-5}
\begin{tabular}{cccc} \hline
Time Bin   & MDP (\%)   & $\Pi_X$ $\pm$ error(\%) & $\Psi_X$ $\pm$ error (deg) \\ \hline
T1 & 11.4      & 30.5 $\pm$ 3.7        & 128.2 $\pm$ 3.5 \\
T2 & 9.9       & 24.0 $\pm$ 3.2        & 131.6 $\pm$ 3.9 \\
T3 & 11.5      & 16.0 $\pm$ 3.8        & 120.5 $\pm$ 6.8 \\
T4 & 10.9      & 14.2 $\pm$ 3.6        & 137.8 $\pm$ 7.2 \\ \hline
\end{tabular}
\end{table}

%%%%%%%%%%%%%%%%%%%%%%%%%%%

\begin{table}
\caption {X-ray polarization variability in PKS 2155$-$304 in 2$-$3 and 3$-$5 keV band.}
\label{table-235}
\centering
\begin{tabular}{ccccccc}
\hline
\multirow{2}{*}{Time Bin} & \multicolumn{3}{c|}{2$-$3 keV} & \multicolumn{3}{c}{3$-$5 keV} \\ \cline{2-7}
 & $MDP$ (\%) & $\Pi_X \pm$ error (\%) & $\Psi_X \pm$ error (deg) 
 & $MDP$ (\%) & $\Pi_X \pm$ error (\%) & $\Psi_X \pm$ error (deg) \\ \hline

T1 & 13 & 26.7 $\pm$ 4.2 & 130 $\pm$ 5 & 14 & 33.2 $\pm$ 4.6 & 130 $\pm$ 4 \\
T2 & 12  & 23.2 $\pm$ 3.8 & 133 $\pm$ 5 & 13 & 27.1 $\pm$ 4.2 & 126 $\pm$ 4 \\
T3 & 13 & 16.6 $\pm$ 4.4 & 122 $\pm$ 8 & 14 & 23.2 $\pm$ 4.7 & 128 $\pm$ 6 \\
T4 & 13 & 19.4 $\pm$ 4.3 & 141 $\pm$ 6 & 14 & 16.7 $\pm$ 4.6 & 125 $\pm$ 8 \\ \hline
\end{tabular}
\end{table}

%%%%%%%%%%%%%%%%%%%%%%%%%

\begin{figure*}
 \centering
\includegraphics[scale=0.3]{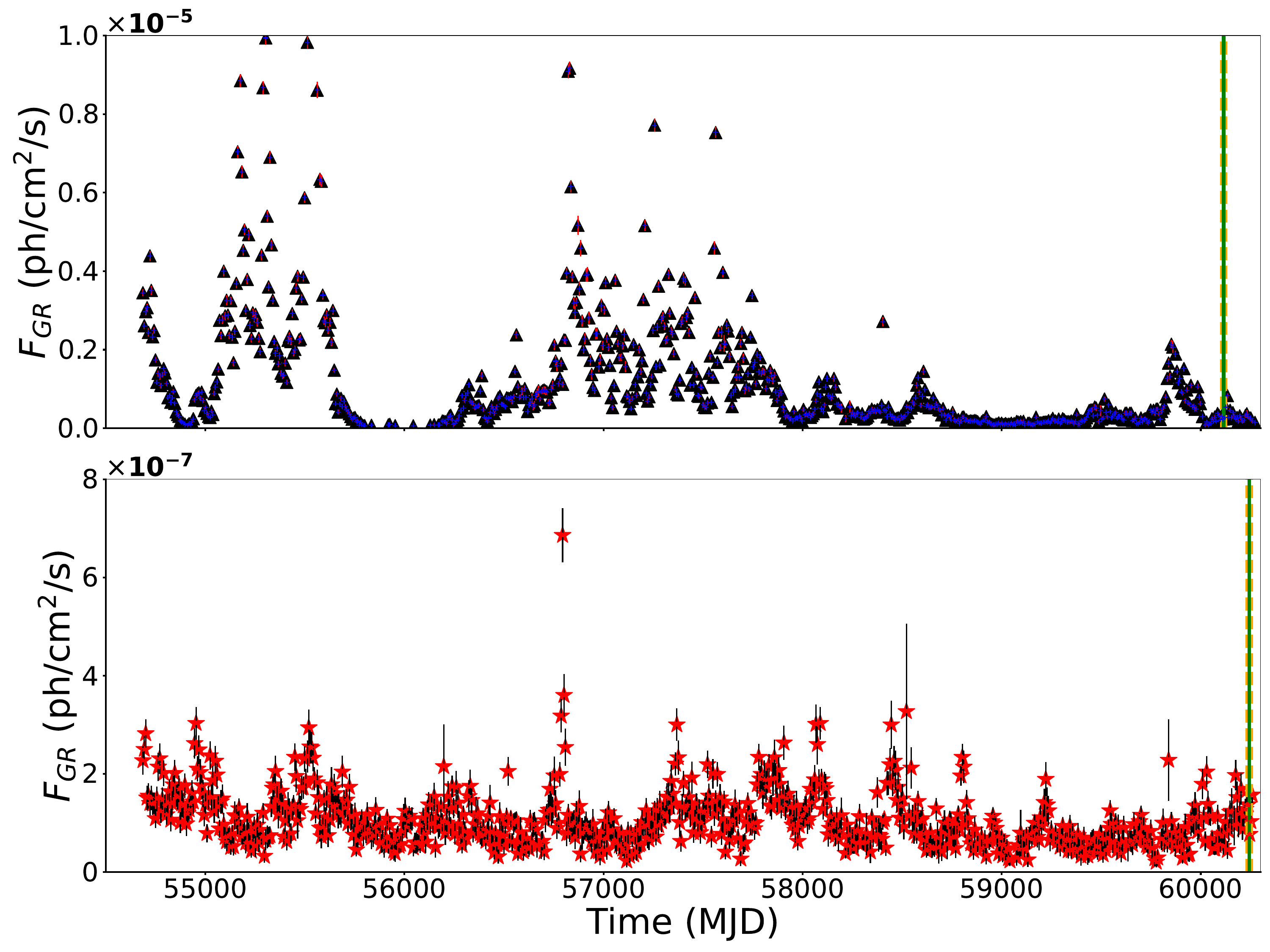}
\caption{The long term $\gamma$-ray light curve in the 100 MeV to 300 GeV band for the sources 3C 454.3 (top panel) and PKS 2155$-$304 (bottom panel). The epoch of \textit{IXPE} 
observation is marked with a green dotted line.}
\label{figure-4}
\end{figure*}

\begin{figure*}
 \centering
\includegraphics[scale=0.25]{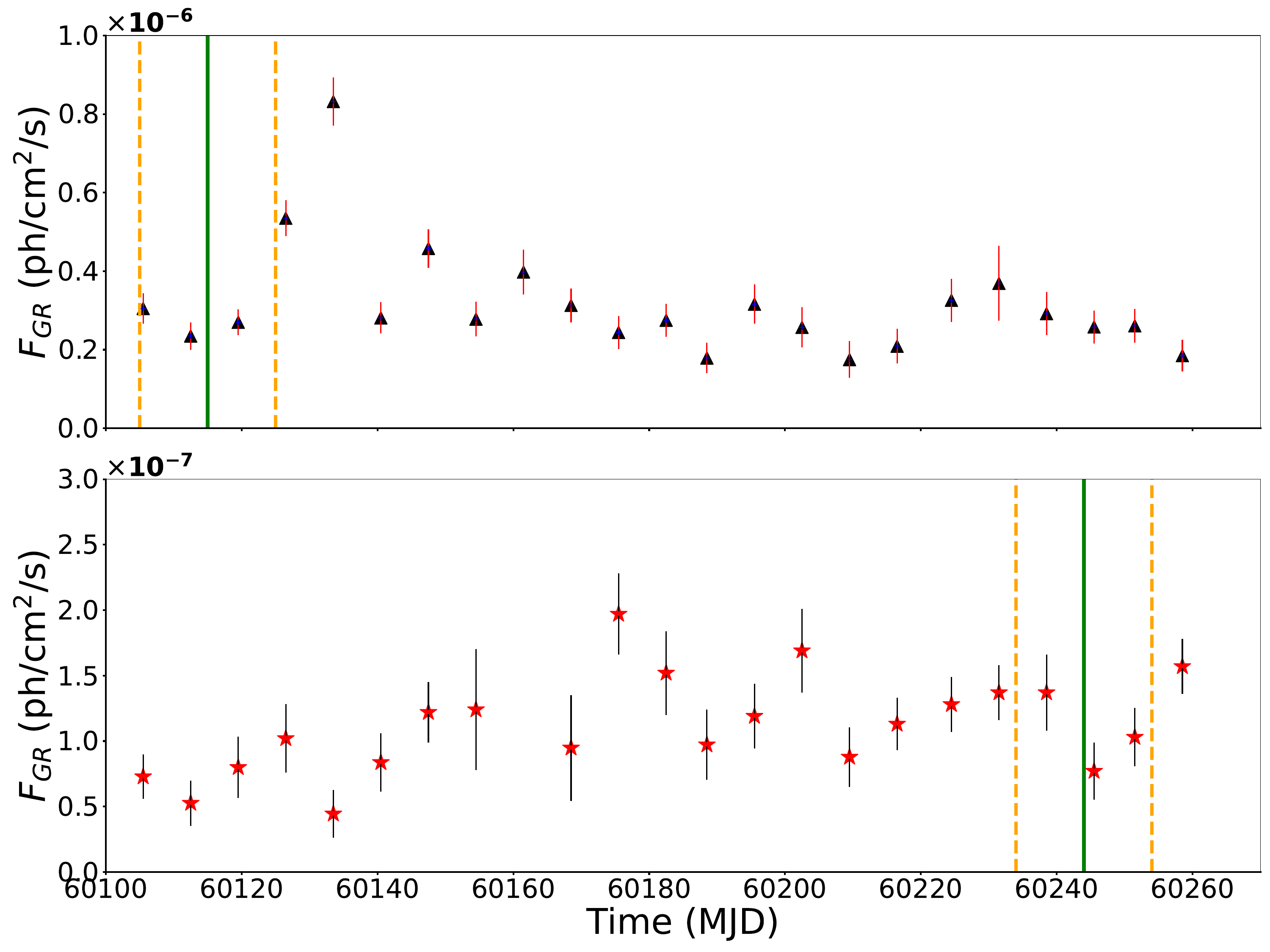}
\caption{The monthly binned $\gamma$-ray light curves spanning a  period of about 250 days for 3C 454.3 (top panel) and 300 days for PKS 2155$-$304 (bottom panel). 
The epoch of \textit{IXPE} observation is shown as a solid line, while the two dotted lines on either side of the \textit{IXPE} pointing
shows the time period used to generate the SEDs.}
\label{figure-5}
\end{figure*}

\subsection{Broadband SED analysis}
\begin{figure*}
 \centering
\includegraphics[scale=0.4, angle=-90]{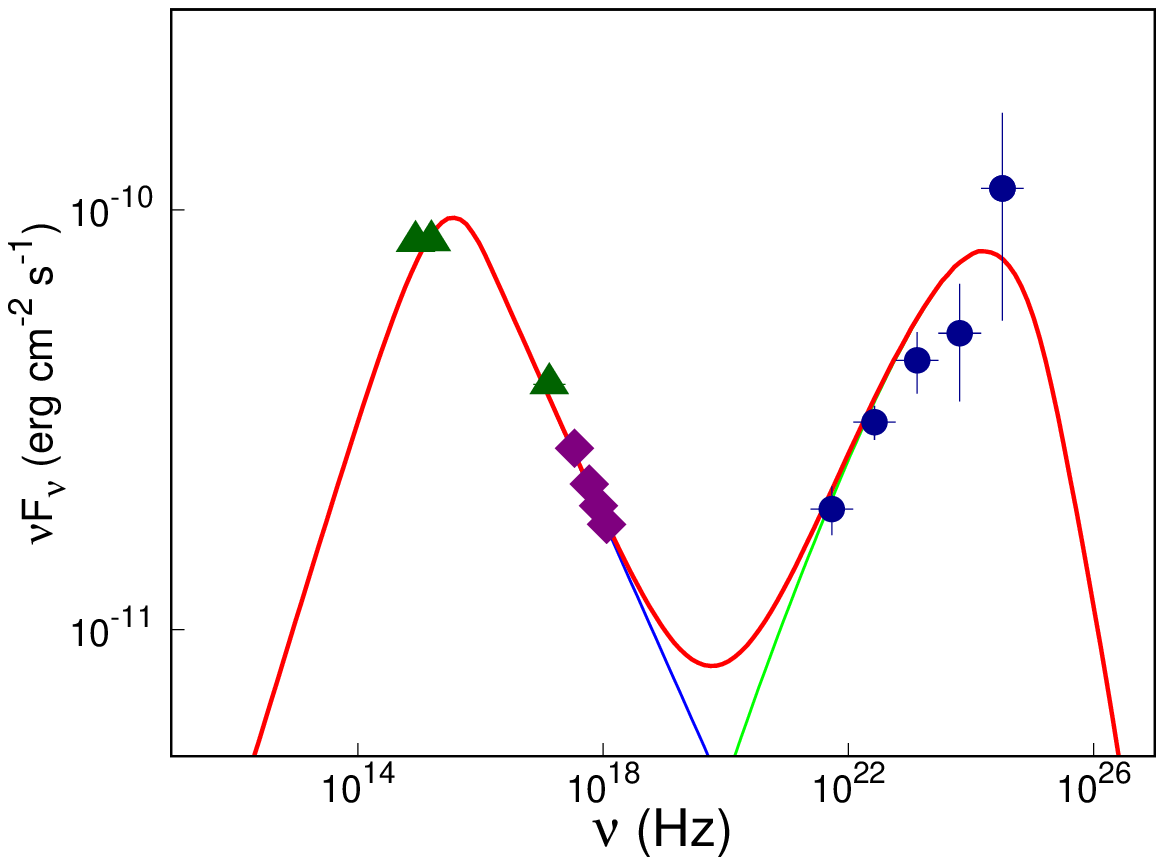}
\includegraphics[scale=0.4, angle=-90]{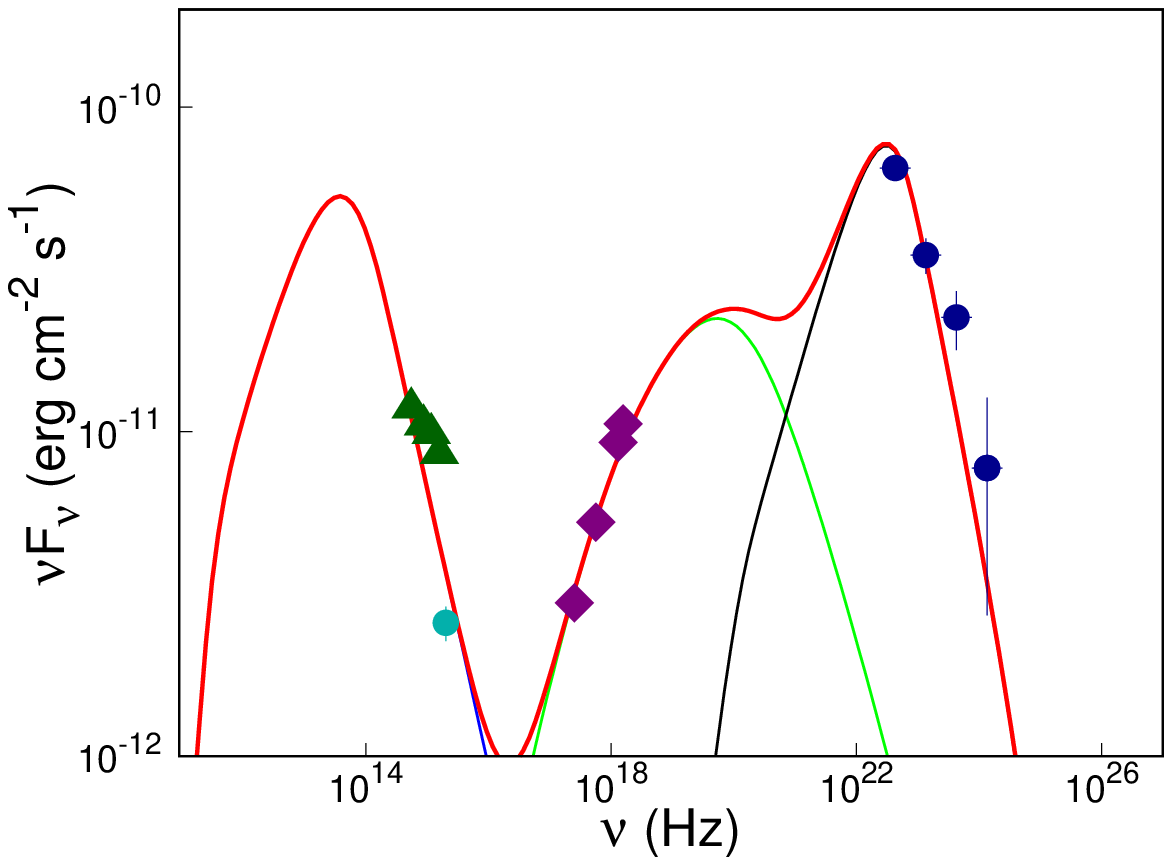}
\caption{The SEDs of PKS 2155$-$304 (left) and 3C 454.3 (right) generated during the 
epochs of \textit{IXPE} observations Here, the blue line represents synchrotron emission, green indicates the emission due to SSC process, black denotes the emission from EC process and the red solid line is the sum of all the emission components. In the SEDs the green triangles are from {\it Swift}-UVOT, the purple diamonds are from {\it Swift}-XRT, the magenta circle is from {\it AstroSat}-UVIT and the blue circles are from {\it Fermi}.}
\label{figure-6}
\end{figure*}

The long term $\gamma$-ray light curve in the energy range of 100 MeV to 300 GeV, for the two sources PKS 2155$-$304 and 3C 454.3 is shown  
in Fig. \ref{figure-4} and it exhibits significant flux variations. The dotted vertical lines correspond to the epochs for which \textit{IXPE} observations are available. A segment of this lightcurve, spanning 300 days for PKS 2155-304 and 250 days for 3C 454.3, which overlays the \textit{IXPE} observations (solid vertical lines) is shown in Fig. \ref{figure-5}.
%A zoomed version of the same figure for a duration of about 250 days is 
%In this figure the region marked in solid lines that encompasses the
%epochs of \textit{IXPE} observations were used to generate the SEDs of the 
%sources. For those selected regions, 
Quasi-simultaneous multi wavelength-data is used to construct the broadband SED
of the sources coincident with the 
%using the data acquired across wavelengths that are quasi-simultaneous with the 
\textit{IXPE} observations (see Table \ref{table-2} and Fig. \ref{figure-6}).
%, we constructed the broadband SEDs for both the sources. 
The observed broadband SEDs are reproduced using an one zone leptonic model \citep{2018RAA....18...35S} where, most of the blazar 
emission is assumed to originate from a spherical blob of size $R$ populated with broken power-law electron distribution given as,
\begin{align} \label{eq:broken}
	N(\gamma)\,d\gamma = \left\{
\begin{array}{ll}
	K\,\gamma^{-p_1}\,d\gamma&\textrm{for}\quad \mbox {~$\gamma_{\rm min}<\gamma<\gamma_b$~} \\
	K\,\gamma_b^{p_2-p_1}\gamma^{-p_2}\,d\gamma&\textrm{for}\quad \mbox {~$\gamma_b<\gamma<\gamma_{\rm max}$~}
\end{array}
\right.
\end{align} 
Here, $\gamma$ is the electron Lorentz factor and, $p_1$ and $p_2$ are the low and high energy power-law indices with $\gamma_b$ the electron Lorentz factor corresponding to the break energy. The emission region is permeated with a magnetic field $B$ and moves down the jet with a bulk Lorentz factor $\Gamma$. The non-thermal electron
distribution lose their energy through synchrotron, SSC, and EC radiative processes. The target ambient photon field for the EC process is chosen to be a blackbody
emission from the dusty environment at temperature 1000 K.
The model parameters are further constrained by imposing equipartition between the magnetic field and particle energy densities.
For the case of PKS 2155$-$304, the broadband
SED can be reproduced by considering only synchrotron and SSC processes; while, to reproduce the broadband SED of 3C 454.3, we had to include EC process also.  
In Fig. \ref{figure-6} we show the model curves along with the observed fluxes and in Table \ref{table-6}, we provide the corresponding model parameters. We fixed $\gamma_{\rm min}$ at 50 and $\gamma_{\rm max}$ at 5.0$\times$10$^{8}$ and 9.3$\times$10$^{5}$ for PKS 2155$-$304 and 3C454.3, respectively. 
%We note that in our modeling, the values of $\gamma_{\rm max}$ were fixed at 5.0$\times$10$^{8}$ and 9.3$\times$10$^{5}$ for PKS 2155$-$304 and 3C454.3. 
These high values were chosen to ensure that the model spectrum extends to significantly large energy. In practice, as evident from Fig. \ref{figure-6} that the synchrotron emission from the electrons above a few $10^{5}$ for PKS 2155$-$304 and above a few $10^{4}$ for 3C 454.3 do not contribute significantly to the observed emission. Therefore, the fits are not sensitive to the choice of $\gamma_{\rm max}$. The values we report should thus be regarded as modeling choices, rather than as physically required extreme electron energies.
The target photon density was 7.6$\times$10$^{-5}$ $erg$ $cm^{-3}$  for 3C 454.3.

%We fixed $\gamma_{\rm min}$ and $\gamma_{\rm max}$ at 50 and 9.3$\times$10$^{5}$  and assumed the viewing angle to be 2°. 

\begin{table}
\centering
\caption{Results of the broadband SED analysis carried out for the epochs of polarization detection. Here, p and q are the particle indices while $\gamma_b$ and B (Gauss) are the break energy and magnetic field. The emission region size (R) is given in log(cm).}
\label{table-6}
\begin{tabular}{ccc}
\hline
Parameter & PKS 2155$-$304  & 3C 454.3  \\
\hline
		$p_1$ & 2.0 & 1.7 \\
		$p_2$ & 3.6 & 4.7 \\
		$\gamma_b$ & 15671.4 & 1027.8 \\
    log (R) (cm) & 17.0 & 16.5 \\
		$\Gamma$ & 11 & 50 \\
		B (G) & 0.12 & 0.84 \\
		\hline
\end{tabular}
\end{table}

The broadband SED model suggests that the X-ray emission from PKS 2155$-$304 is due to synchrotron process and hence it will be polarized whereas the SSC emission dominates the X-ray energy band of 3C 454.3, resulting in less/negligible polarization. For PKS 2155$-$304, we can estimate the degree of polarization from the synchrotron process using the following equation from \cite{1986rpa..book.....R},
\begin{equation}
\label{eq:PD2}
    \Pi_{X}=\frac{p + 1}{p + 7/3}
\end{equation}
where, p is the particle index. From the photon index $s$ obtained from the spectro-polarimetric fit to the \textit{IXPE} observations (see Table \ref{table-4}), we calculated the particle index, as p = 2$\alpha$ + 1 (where $\alpha$ $-$1  = $s$), for which we obtained a value of $\Pi_X$ = 88\%, for aligned magnetic field. However, \textit{IXPE} observations suggest $\Pi_X$ = (20.9 $\pm$ 1.8)\%, much lower than the expected value. A plausible reason could be that the magnetic field lines in the emission region are significantly tangled, thereby reducing the net observed polarization. The large discrepancy between the predicted polarization for an aligned magnetic field and the observed \textit{IXPE} value can be quantitatively explained by considering a multi-zone turbulent magnetic field structure, where the net polarization decreases as $\Pi_{o b s}=\Pi_{\max } N^{-1 / 2}$, where $N$ number emission zones with randomly oriented fields \citep{1966MNRAS.133...67B}. Using this, a reduction by a factor of $\sim$ 4 would correspond to roughly 16 independent turbulent zones, which is consistent with models of magnetic turbulence in blazar jets \citep{2014ApJ...780...87M, 2014ApJ...789...66Z, 2016ApJ...817...63Z}.

\section{Discussion}
\label{sec:dis}

The observed X-rays in blazars can be produced by different physical processes under the leptonic emission scenario such as synchrotron emission from the relativistic electron population in the jet or the inverse Compton scattering of synchrotron photons/external photon field. Alternatively, they can also be produced via synchrotron emission from a relativistic proton population or nuclear cascades. One way to determine the dominant X-ray emission mechanism in blazars is via broadband SED modeling. Unlike previous studies that focused solely on \textit{IXPE} data, we integrate information from \textit{Fermi}, \textit{Swift}, and \textit{AstroSat}, constructing a more comprehensive picture of the emission mechanisms. The SED modeling for these specific \textit{IXPE} epochs has not been presented before, allowing us to place additional constraints on the physical conditions in the jet. The conclusion drawn from the SED modeling can be further scrutinized by measuring the X-ray polarization which was made possible with the advent of \textit{IXPE} since the beginning of the year 2022. This is because different X-ray emission processes have inherently different X-ray polarization characteristics. For instance, synchrotron radiation, as established by \cite{1959ApJ...130..241W}, is inherently polarized, while the synchrotron self Compton process generally produces less polarized emission than synchrotron process \citep{2019ApJ...885...76P, 2024ApJ...967...93Z}. In this work, we analyzed \textit{IXPE} observations as well as carried out broadband SED modeling of one HSP blazar PKS 2155$-$304 and one LSP blazar 3C 454.3 to test the X-ray emission mechanisms in them. 

\begin{figure}
 \centering
\includegraphics[scale=0.3]{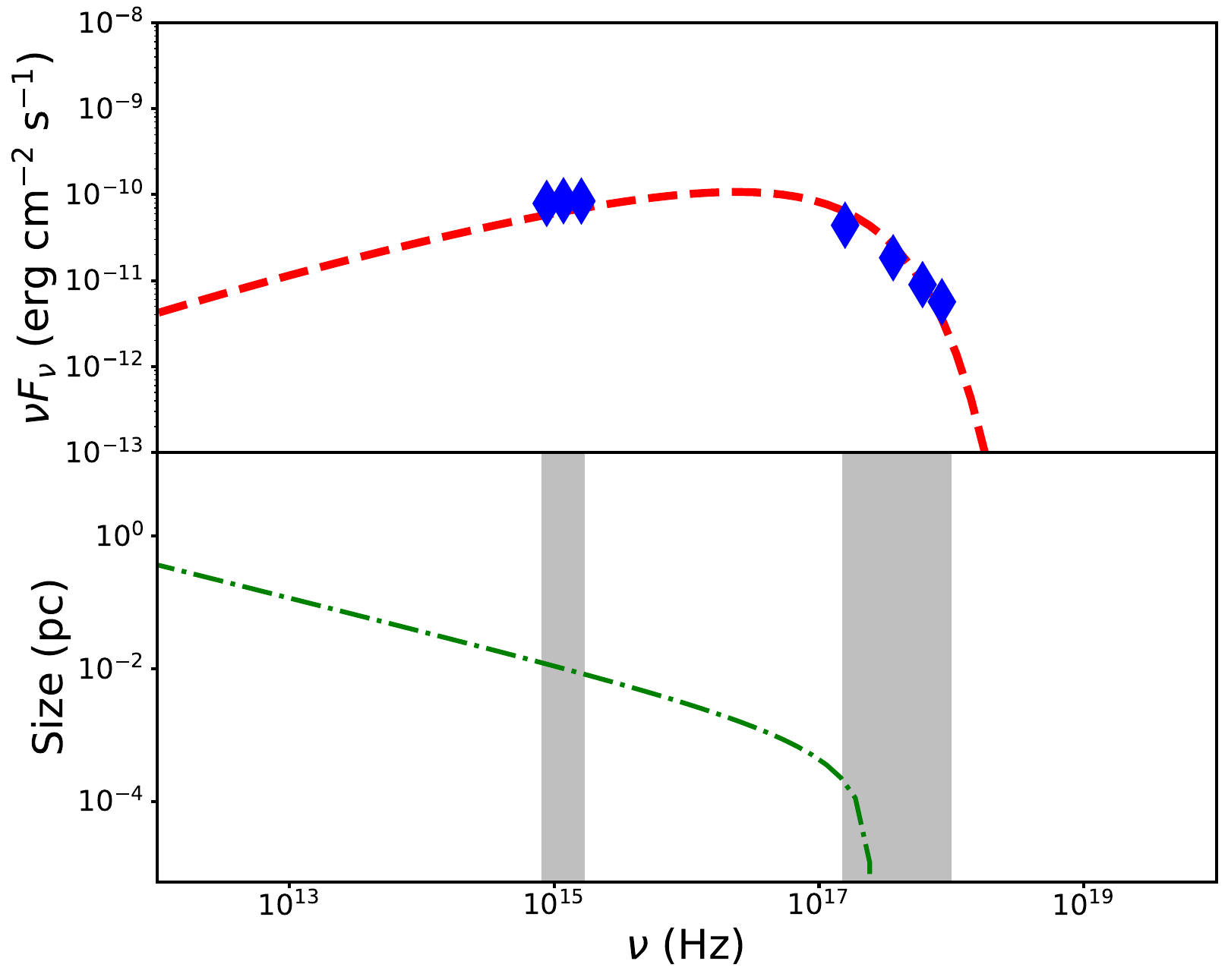}
\caption{The figure consists of two subplots: The top subplot shows the synchrotron spectrum due to a stratified jet model  \citep{2024ApJ...975..185B} along with the observed fluxes. The bottom panel  indicates the spatial extension of their respective emission regions. The dotted line (top panel) is the fit to the observed data
points in the optical and X-ray bands. The shaded regions (in
the bottom panel) correspond to the optical and X-ray energies}
\label{figure-7}
\end{figure}

For the source PKS 2155$-$304, in the 2$-$8 keV band from model independent PCUBE analysis we found $\Pi_X$ = (20.9 $\pm$ 1.8)\% and $\psi_X$ = (130 $\pm$ 2.5) deg, which is in agreement with model dependent spectro-polarimetric analysis. In addition to measuring a high degree of X-ray polarization, we also found $\Pi_X$ to vary both with energy and time. Our division of the dataset into finer time bins allows a more granular view of polarization variability. We found a high value of $\Pi_X$ = (25.3 $\pm$ 2.3)\% in the 3$-$5 keV band and a low value of $\Pi_X$ = (20.9 $\pm$ 2.1)\% in the 2$-$3 keV band. During the same epoch, the optical polarization was found to be much lower with a value of (4.6 $\pm$ 0.4)\% in the B-band \citep{2024A&A...689A.119K}. There is thus clear evidence for energy dependent polarization. These results strengthen the case for a structured jet model where X-ray emission originates from a compact acceleration zone near the shock front, while lower-energy optical emission is produced in a broader, more turbulent region. These observations support that the emission region may be stratified with the X-ray emission arising in the vicinity of the particle acceleration site (shock front) while the optical emission may have contribution from the extended region \citep{2024ApJ...975..185B}. This aligns with models of shock acceleration in blazar jets \citep{1980MNRAS.193..439L, 1985ApJ...298..114M, 1991bja..book.....H} but provides additional observational constraints based on polarization evolution. To explore this further, we considered a scenario (similar to the one presented in \citealt{2024ApJ...975..185B}) where, the emission region is situated around a shock front. Electrons are accelerated near the shock front and are advected into the jet medium, while losing energy through radiative processes. Due to shock compression, the magnetic field lines near the shock front are expected to be aligned. Whereas, they may be tangled at farther regions due to the turbulent nature of the jet. Consequently, synchrotron emission occurring near the shock front will be more polarized compared to extended regions. From the SED modeling (Fig. \ref{figure-7}) of PKS 2155$-$304, we found that the X-ray emission involves higher electron energies than the optical emission. Using the model parameters, we found the electron Lorentz factors responsible for optical and X-ray emission as $\gamma_o$ =12018 and $\gamma_X$ =276363. If we assume they advect at relativistic velocity then from the synchrotron cooling time scale for the X-ray and optical emitting electrons, the extension of the optical and the X-ray emission regions can be estimated as 9.27$\times$10$^{-3}$ pc and 4.03$\times$10$^{-4}$ pc. In Fig. \ref{figure-7}, we show the synchrotron spectrum from the stratified emission region along with the observed X-ray and optical fluxes. In the bottom panel we show the extension of the emission region for different electron/photon energies. While shocks provide a natural explanation for enhanced X-ray polarization, it is important to note that this is not the only viable scenario. As highlighted by \cite{2024ApJ...967...93Z}, if particles are accelerated in a compact region and subsequently diffuse or advect into a larger zone with a partially ordered magnetic field, the X-ray polarization can exceed that in the optical band independent of the underlying acceleration mechanism, such as shocks, magnetic reconnection, or turbulence.

For 3C 454.3, we could not detect appreciable X-ray polarization in the 2$-$8 keV band. We obtained a value of $\Pi_X$ = (3.7 $\pm$ 4.3)\%, much lower than the MDP of 13\%. This non-detection therefore favours inverse Compton mechanism for the production of X-rays as a plausible mechanism. Another viable emission mechanism involving hadronic process can be nuclear cascades. Nevertheless, the broadband SED generated for both the sources during the epoch of \textit{IXPE} observations using quasi-simultaneous data acquired at other wavelengths were well explained by leptonic models. The confirmation of an SSC+EC scenario for 3C 454.3 is consistent with past studies \citep{2010ApJ...714L.303F, 2017MNRAS.470.3283S, 2022Univ....8..585F} but provides additional constraints by considering new observational epochs. Similar to the source PKS 2155$-$304, significant detection of X-ray polarization was detected for other HSP blazars Mrk 421 \citep{2022ApJ...938L...7D} and Mrk 501 \citep{2022Natur.611..677L} using \textit{IXPE} observations.

\section{Summary}
\label{sec:con}
We carried out an investigation of the X-ray emission mechanism in the HSP blazar  PKS 2155$-$304 and in the LSP blazar 3C 454.3 via X-ray polarization observations with \textit{IXPE} and broadband SED modeling. The results are summarized below.
\begin{enumerate}
\item In PKS 2155$-$304, we observed a high degree of polarization with $\Pi_X$ = (20.9 $\pm$ 1.8)\% and $\Psi_X$ = (130 $\pm$ 2.5) deg in the 2$-$8 keV band from model independent analysis. This is also in agreement with that found from spectro-polarimetric analysis with $\Pi_X$ = (23.3 $\pm$ 2.5)\% and $\Psi_X$ = (129 $\pm$ 3.1) deg.  The high value of X-ray polarization found here is also in agreement with
that reported by \cite{2024ApJ...963L..41H}.

\item We found $\Pi_X$ to vary both in energy and time. In the 3$-$5 keV band we found $\Pi_X$ = (25.1 $\pm$ 2.3)\%, while in the 2$-$3 keV band we found $\Pi_X$ = (20.9 $\pm$ 2.1)\%. Though $\Pi_X$ in the 2$-$3 keV and 3$-$5 keV bands agree within errors (given the large error bars), there is a tendency of $\Pi_X$ in the 3$-$5 keV band to be larger than that in the 2$-$3 keV band. There is thus an indication of energy dependent polarization.

\item Broadband SEDs generated using quasi-simultaneous data  during the epochs of \textit{IXPE} observations for PKS 2155$-$304 and 3C 454.3 are well explained by leptonic processes.

\item 
The lower optical polarization compared to X-ray indicates that electrons are primarily accelerated at the shock front. High-energy X-rays are produced close to this acceleration zone, while optical emission originates from regions farther downstream in the jet. SED modeling under this scenario suggests that the magnetic field lines become disordered over a characteristic length scale of approximately $\sim$0.01pc from the shock front.

\end{enumerate}

Our observations support a scenario of X-rays being produced via synchrotron process from relativistic electrons in the jet of the HSP blazar PKS 2155$-$304. However, in the case of LSP balzar 3C 454.3,  our analysis favour inverse Compton emission mechanism for the production of X-rays.

This study's findings contribute to expanding the list of blazars with detected X-ray polarization from {\it IXPE} observations. This growing dataset provides a valuable opportunity to systematically investigate the polarization properties of blazars in the X-ray regime, which in turn can offer crucial insights into the particle acceleration mechanisms and magnetic field structures within their relativistic jets. However, to achieve a more complete understanding of the high-energy emission processes in blazars, it is essential to conduct further X-ray polarimetric observations of additional sources. These observations should ideally be complemented by simultaneous multi-wavelength monitoring across optical, UV, and radio bands. Such coordinated observations can help establish correlations between the polarization properties at different wavelengths, thereby constraining the emission mechanisms responsible for the observed radiation. For instance, comparing X-ray and optical polarization can reveal whether the high-energy emission is dominated by SSC or hadronic processes, while radio polarization studies can shed light on the large-scale magnetic field configurations in blazar jets.

Moreover, increasing the sample of blazars with X-ray polarization measurements can provide a statistical perspective on polarization variability across different subclasses of blazars, such as HSP, ISP, and LSP sources. Understanding whether certain classes exhibit systematically higher polarization degrees, or whether polarization variability is linked to changes in flux states, can refine our theoretical models of blazar jet emission.
\begin{table*}[htb]
\centering
\caption{The log of X-ray polarization observations by \textit{IXPE} till August 2025. The source, date of observation and the measured polarization is given.}
\label{table-ixpetot}
\begin{tabular}{c c c c }
			\hline
			  Source & OBS date & $\Pi_{X} (\%$) &  Reference\\
			\hline 
			Mrk 421 & 2022-05-04 & 15$\pm$2  &  \cite{2022ApJ...938L...7D} \\
           & 2022-06-04 & 10$\pm$1 & \cite{2023NatAs...7.1245D} \\
           & 2022-06-07 & 10$\pm$1 & \cite{2023NatAs...7.1245D}  \\
           & 2022-12-06 & 14$\pm$1 & \cite{2024AA...681A..12K} \\
           & 2023-12-06 & 8.5$\pm$0.5 & \cite{2024ApJ...975..185B} \\
            \hline 
            Mrk 501 & 2022-03-08 & 10$\pm$2  & \cite{2022Natur.611..677L} \\
           & 2022-03-27 & 11$\pm$2 & \cite{2022Natur.611..677L} \\
           & 2022-07-09 & 5.8$\pm$2.3 & \cite{2024arXiv240211949H} \\
           & 2023-02-12 & 10$\pm$3.2 & \cite{2024arXiv240211949H} \\
           & 2023-03-19 & 6.1$\pm$2.8 & \cite{2024arXiv240211949H} \\
         & 2023-04-16 & 15.8$\pm$2.8 & \cite{2024arXiv240211949H} \\
        
         \hline
         1ES 1959+650 & 2022-05-03 & 8.0 $\pm$ 2.3  & \cite{2024ApJ...963....5E} \\
          & 2022-06-09 &  $<$5.1\% & \cite{2024ApJ...963....5E} \\
           & 2022-10-28 & 9.0 $\pm$ 1.6\% & \cite{2024JApA...45...35B} \\
           & 2023-08-14 & 12.5 $\pm$ 0.7\% &  \cite{2024JApA...45...35B} \\
           \hline
      1ES 0229+200 & 2023-01-15  & 17.9\% $\pm$ 2.8\%  &  \cite{2023ApJ...959...61E}\\
        \hline
      PG1553+113 & 2023-01-15  & 10\% $\pm$ 2\%  &  \cite{2023ApJ...953L..28M}\\
            \hline
      PKS 2155$-$304 & 2022-03-08 &  21.9\% $\pm$ 1.9\% & \cite{2024ApJ...963L..41H} \\ 
        \hline
       H 1426+428 & 2024-07-05 & 19.3\% $\pm$ 4.5\%   & \cite{2025arXiv250414869H} \\
        & 2024-05-27 &  $<$19.5\%. & \cite{2025arXiv250414869H} \\
        \hline

		\end{tabular}
\end{table*}

%\item Hence, it disfavors the magnetic re-connection and stochastic mechanism of particle accelerations.

\section*{Acknowledgments}
The Imaging X-ray Polarimetry Explorer (\textit{IXPE}) is a joint US and Italian mission. The US contribution is supported by the National Aeronautics and Space Administration (NASA) and led and managed by its Marshall Space Flight  Center (MSFC), with industry partner Ball Aerospace (contract NNM15AA18C). The Italian contribution is supported by the Italian Space Agency (Agenzia Spaziale Italiana, ASI) through contract ASI-OHBI-2017-12-I.0, agreements ASI-INAF-2017-12-H0 and ASI-INFN-2017.13-H0, and its Space Science Data Center (SSDC) with agreements ASI- INAF-2022-14- HH.0 and ASI-INFN 2021-43-HH.0, and by the Istituto Nazionale di Astrofisica (INAF) and the Istituto Nazionale di Fisica Nucleare (INFN) in Italy. This research used data products provided by the {\it IXPE} Team (MSFC, SSDC, INAF, and INFN) and distributed with additional software tools by the High-Energy Astrophysics Science Archive Research Center (HEASARC), at NASA Goddard Space Flight Center (GSFC). Athira M Bharathan acknowledges the Department of Science and Technology (DST) for the INSPIRE FELLOWSHIP (IF200255). Also thank the Center for Research, CHRIST (Deemed to be University) for all their support during the course of this work. 

%%%%%%%%%%%%%%%%%%%%%%%%%%%%%%%%%%%%%%%%%%%%%%%%%%
\section*{Data Availability}

All the data used here for analysis is publicly available and the results are incorporated in the paper.



\bibliographystyle{harv}  
\bibliography{main}

\bio{}
\endbio

\bio{}
\endbio

\end{document}